\documentstyle[a4,11pt] {article}
\begin{document}

\setlength{\baselineskip}{20pt}

\noindent
{\Large\bf Response to the comments of Dwivedi and Srivastava on the 
propagation and dissipation of Alfv\'en waves in coronal holes}

\vspace{4mm}

\setlength{\baselineskip}{15pt}

\noindent
{\bf Suresh Chandra and B.K. Kumthekar}

\vspace{2mm}

\noindent
School of Physics, S.M.V.D. University, Katra 182 320, (J\&K)

\vspace{2mm}

\noindent
Email: suresh492000@yahoo.co.in

\vspace{6mm}

\noindent
{\bf Abstract:} Chandra [1] made an attempt to show that the work of Dwivedi and
 Srivastava [2] (hereinafter DS) can be investigated even analytically and their
 results are erroneous. Dwivedi and Srivastava [3] picked up  some values of Chandra 
[1] and tried to show that they are not physically acceptable. Some results 
of Chandra [1] are not physically acceptable, as these are the outcome of 
the wrong approach of DS. However, the results are numerically correct 
whereas the results of DS are numerically wrong. 

\vspace{2mm}


\section{Introduction}

In order to show that the work of Dwivedi and Srivastava [2] (hereinafter DS) 
can be investigated even analytically,  Chandra [1] used the condition $\omega 
\nu > > v_A^2$ for the case of $\eta = 0$ and $\omega \eta < < v_A^2$ for the 
case of $\nu = 0$. His also proved that the results of DS are wrong. Dwivedi and
 Srivastava [3] should not misguide the scientific
community by wrongly interpreting these conditions, as the numerical 
calculations can also prove that the results of DS are wrong. Moreover, as a
consequence of the wrong approach of DS, some numerical results obtained by
Chandra [1]  are not physically acceptable. In the present work we have shown 
numerically that the results of DS are wrong.

\section{Formulation}

The dispersion relation derived by Pek\"unl\"u et al. [4] for the MHD 
equations can be expressed as [1, 2, 4]
\begin{eqnarray}
\nu \eta k^4 + [v_A^2  - i \omega (\nu + \eta)] k^2 - \omega^2 =  0 
\label{eq0}
\end{eqnarray}

\noindent
Consideration of an atmosphere with (i) $\nu = 0$ (no viscosity) or (ii) with 
$\eta = 0$ (no magnetic diffusivity) may not be a physically acceptable 
situation. However, DS discussed these two cases numerically, where they got 
wrong results.  Let us now investigate these
 two cases numerically. For either of the cases, the first term $\nu \eta k^4$ 
is zero and we have
\begin{eqnarray}
\Big[v_A^2  - i \omega (\nu + \eta)\Big] k^2 - \omega^2 =  0 \label{eq1}
\end{eqnarray}

\noindent
It is matter of surprise that Dwivedi and Srivastava [3] are asking for
solving equation (ref{eq0}) numerically.
For real values of $\omega$, the wave number $k$ can be assumed to have complex
 values. For $k = k_r + i k_i$, we can derive the values of $k_r$ and $k_i$ 
for the said two cases as the following:

\vspace{2mm}

(i) There is viscosity only . That is, there is no magnetic diffusivity
($\eta = 0$).  For this case, equation (\ref{eq1}) gives [1, 2]
\begin{eqnarray}
k_r^4 - P \ k_r^2 - Q^2 = 0 \hspace{1.5cm} \mbox{and} \hspace{1.5cm} k_r k_i =
Q  \label{eq2}
\end{eqnarray}

\noindent
where
\begin{eqnarray}
P = \frac{\omega^2 v_A^2}{v_A^4 + \omega^2 \nu^2}  \hspace{1.5cm} \mbox{and}
\hspace{1.5cm} Q = \frac{\omega^3 \nu}{2(v_A^4 + \omega^2 \nu^2)}
\nonumber 
\end{eqnarray}

\noindent
Form equation (\ref{eq2}), we get
\begin{eqnarray}
k_r^2 = \frac{\omega^2 v_A^2}{2 (v_A^4 + \omega^2 \nu^2)} \Big[1 + \sqrt{
1 + (\omega \nu/v_A^2)^2}\Big] \label{eq3}
\end{eqnarray}

(ii) There is magnetic diffusivity only. That is, there is no viscosity
($\nu = 0$). For this case, equation (\ref{eq1}) gives [1, 2]
\begin{eqnarray}
k_r^4 - P'k_r^2 - Q'^2 = 0 \hspace{1.5cm} \mbox{and} \hspace{1.5cm} k_r k_i = Q'
 \label{eq4}
\end{eqnarray}

\noindent
where
\begin{eqnarray}
P' = \frac{\omega^2 v_A^2}{v_A^4 + \omega^2 \eta^2}  \hspace{1.5cm} \mbox{and}
 \hspace{1.5cm} Q' = \frac{\omega^3 \eta}{2(v_A^4 + \omega^2 \eta^2)}
\nonumber 
\end{eqnarray}

\noindent
Form equation (\ref{eq4}), we get
\begin{eqnarray}
k_r^2 = \frac{\omega^2 \ v_A^2}{2 (v_A^4 + \omega^2 \eta^2)} \Big[1 + \sqrt{
1 + (\omega \eta/v_A^2)^2}\Big] \label{eq5}
\end{eqnarray}

\section{Results}

In order to get numerical values for $k_r$ and $k_i$, for the two cases, we
require values of physical parameters. The values of physical parameters are
provided by Chandra [1] and some of them are given in Table 1. In order to find 
out the source of error in the work of DS, we asked them several times to 
provide us the values of physical parameters used in their work, but they never 
showed the courtesy to reply. For $\tau = 10^{-4}$ s, the 
calculations are made and the results are given in Table 1.
The damping length scale $D$ and the wavelength $\lambda$ of the wave are
\begin{eqnarray}
D = \frac{2 \pi}{k_i} \hspace{1.5cm} \mbox{and} \hspace{1.5cm} \lambda = 
\frac{2 \pi}{k_r} \nonumber
\end{eqnarray}

\noindent
The results for the case $\eta = 0$ show that the damping length scale and the 
wavelength are of the same magnitude, which is of the order of $10^4$ m. For the
 case $\nu = 0$, the damping length scale comes out of the order of 10$^{11}$ m.This physically unacceptable result is obviously a consequence of the wrong
 approach of DS. Comparison of the present results with those of DS show that
their results are wrong. Even the nature of the variation of $D$ is quite
different.

\vspace{6mm}

\begin{tabular}{cccc|ll|ll}
\multicolumn{8}{l}{Table 1: Values of $k_r$ and $k_i$ for the two cases.}\\
\hline
 & & & & \multicolumn{2}{c}{$\nu =0$} & \multicolumn{2}{|c}{$\eta =0$}  \\
\cline{5-6}
\cline{7-8}
$R^a$ & $\nu^b$ & $\eta^c$ & $v_A^d$ & $k_r$ & $k_i$ & $k_r$ & $k_i$ \\
\hline
 1.05 &   0.20 &   3.79 &   3.25 &   1.94E-02 &   2.19E-10 &   1.26E-03 &   1.25E-03\\
 1.07 &   0.48 &   2.45 &   3.26 &   1.93E-02 &   1.40E-10 &   8.09E-04 &   8.06E-04\\
 1.09 &   0.93 &   1.80 &   3.26 &   1.93E-02 &   1.03E-10 &   5.82E-04 &   5.81E-04\\
 1.11 &   1.56 &   1.44 &   3.26 &   1.93E-02 &   8.23E-11 &   4.48E-04 &   4.48E-04\\
 1.13 &   2.40 &   1.21 &   3.24 &   1.94E-02 &   7.01E-11 &   3.62E-04 &   3.62E-04\\
 1.15 &   3.44 &   1.06 &   3.23 &   1.95E-02 &   6.24E-11 &   3.02E-04 &   3.02E-04\\
 1.17 &   4.65 &   0.96 &   3.21 &   1.96E-02 &   5.77E-11 &   2.60E-04 &   2.60E-04\\
 1.19 &   6.01 &   0.89 &   3.18 &   1.97E-02 &   5.48E-11 &   2.29E-04 &   2.29E-04\\
 1.21 &   7.46 &   0.85 &   3.16 &   1.99E-02 &   5.34E-11 &   2.05E-04 &   2.05E-04\\
 1.23 &   8.92 &   0.83 &   3.13 &   2.00E-02 &   5.30E-11 &   1.88E-04 &   1.88E-04\\
 1.25 &  10.31 &   0.82 &   3.11 &   2.02E-02 &   5.35E-11 &   1.75E-04 &   1.75E-04\\
 1.27 &  11.50 &   0.83 &   3.10 &   2.03E-02 &   5.46E-11 &   1.65E-04 &   1.65E-04\\
 1.29 &  12.40 &   0.85 &   3.10 &   2.03E-02 &   5.63E-11 &   1.59E-04 &   1.59E-04\\
 1.31 &  12.88 &   0.89 &   3.11 &   2.02E-02 &   5.86E-11 &   1.56E-04 &   1.56E-04\\
 1.33 &  12.86 &   0.96 &   3.14 &   2.00E-02 &   6.12E-11 &   1.56E-04 &   1.56E-04\\
 1.35 &  12.26 &   1.06 &   3.20 &   1.97E-02 &   6.43E-11 &   1.60E-04 &   1.60E-04\\
\hline
\multicolumn{8}{l}{\small $^a$Distance from the center of sun in $R_\odot$}\\
\multicolumn{8}{l}{\small $^b$Coefficient of viscosity in 10$^{11}$ m$^2$
s$^{-1}$;}\\
\multicolumn{8}{l}{\small $^c$Magnetic diffusivity in m$^2$ s$^{-1}$}\\
\multicolumn{8}{l}{\small $^d$Alfv\'{e}n velocity in 10$^6$ m s$^{-1}$}\\
\end{tabular}

\vspace{6mm}

\section*{Acknowledgments}
We are thankful to the Department of Science \& Technology (DST), New Delhi and
the Indian Space Research Organization (ISRO), Bangalore for financial support
in the form of research projects. 

\vspace{6mm}

\noindent
{\large\bf  References}
\begin{description}

\item{} [1] Suresh Chandra. Comment on propagation and dissipation of
Alfv\'{e}n waves in  coronal holes. Open Astron J. 2009; {\bf 2}: 16 - 18

\item{} [2] Dwivedi B N and Srivastava A K. On the propagation and dissipation
of Alfv\'{e}n waves in coronal holes. Solar Phys. 2006; {\bf 237}: 143 - 152

\item{} [3] Dwivedi B N and Srivastava A K. Comment on ``Comment on propagation
 and dissipation of Alfv\'{e}n waves in  coronal holes" by Suresh Chandra. Open
 Astron J. 2009; {\bf 2}: 72 - 73 (DS).

\item{} [4] Pek\"unl\"u E R, Bozkurt Z, Afsar M, Soydugan E and Soydugan F.
 Alfv\'{e}n waves in the inner polar coronal hole. Mon. Notices Roy. Astron.
Soc. 2002; {\bf 336}: 1195 - 2000.

\end{description}

\end{document}